\documentstyle[buckow,amssymb]{article}
\newcommand{\g}{\mathcal{G}}

\begin {document}

\pagestyle{empty}
\begin{flushright}
{CERN-TH/2000-381}\\
\end{flushright}
\vspace*{5mm}
\begin{center}
{\bf Space-time symmetries and simple superalgebras} \\
\vspace*{1cm} 
{\bf S. Ferrara} \\
\vspace{0.3cm}

Theoretical Physics Division, CERN \\
CH - 1211 Geneva 23 \\
and\\
Laboratori Nazionali di Frascati, INFN, Italy\\
\vspace*{3.5cm}  
{\bf ABSTRACT} \\ \end{center}
\vspace*{5mm}
\noindent
We describe spinors in Minkowskian spaces with arbitrary signature and their
role in the classification of space-time superalgebras and their R-symmetries in
any dimension.
\vspace*{4cm}
\begin{center} 
{\it Contribution to the Proceedings of the First Workshop of the RTN Network\\
``The quantum structure of spacetime and the geometric nature of fundamental
interactions"\\
Berlin, Germany, 4-10 October 2000}
\end{center}
\vspace*{3.5cm}

\begin{flushleft} CERN-TH/2000-381 \\
December 2000
\end{flushleft}
\vfill\eject

\setcounter{page}{1}
\pagestyle{plain}

\section{Introduction}

We consider  supersymmetry algebras in space-times with arbitrary
signature and minimal number of spinor generators. The
interrelation between super Poincar\'e and superconformal
algebras is elucidated\footnote{The content of this report is based on
Refs.~\cite{aaa} and \cite{bb}}. Minimal superconformal algebras are seen
to have as bosonic part a classical semisimple algebra naturally
associated to the spin group. This algebra, the
Spin$(s,t)$-algebra\cite{aaa}, depends  both on the dimension and on the
signature of space time. We also consider superconformal
algebras, which are classified by the  orthosymplectic algebras.

We then generalize the classification to $N$-extended space-time superalgebras and
notice that R-symmetries may become non-compact depending on the space-time
signature\cite{bb}.  The latter applies to the case of Euclidean super Yang-Mills
theories in four dimensions.

\section{Properties of spinors of SO($\mathbf{V}$)\label{seclif}}

Let $V$ be a real vector space of dimension $D=s+t$ and
$\{v_\mu\}$ a basis of it. On $V$ there is a non degenerate
symmetric bilinear form which in the basis is given by the matrix
 $$\eta_{\mu\nu}={\rm diag}(+,\dots (s \;{\rm times})\dots,
+,-,\dots (t \;{\rm times})\dots, -).$$

We consider the group Spin($V$), the unique double covering of the
connected component of ${\rm SO}(s,t)$ and its spinor
representations. A spinor representation of Spin$(V)^{\Bbb C}$ is an
irreducible complex representation whose highest weights are the
fundamental weights corresponding to the right  extreme nodes in
the Dynkin diagram. These do not descend to representations of
SO$(V)$. A spinor type representation is any irreducible
representation that doesn't descend to SO$(V)$. A spinor
representation of Spin$(V)$ over the reals  is an irreducible
representation over the reals whose complexification is a direct
sum of spin representations\cite{ch,bou,cb,de}.

 Two parameters, the signature $\rho$
mod(8) and the dimension $D$ mod(8) classify the properties of the
 spinor representation. Through this paper we will use the
following notation, $$ \rho=s-t=\rho_0 +8n,\qquad D=s+t=D_0 +8p,$$
where $\rho_0, D_0= 0,\dots 7$. We set $m=p-n,$ so
\begin{eqnarray*}s&=&\frac{1}{2}(D +\rho)=\frac{1}{2}(\rho_0
+D_0)+8n+4m,\\
t&=&\frac{1}{2}(D-\rho)=\frac{1}{2}(D_0-\rho_0)+4m.\end{eqnarray*}

The signature  $\rho$ mod(8) determines if the spinor
representations are real ($\Bbb R$), quaternionic  ($\Bbb H$) or complex
($\Bbb C$) type.  Also note that reality properties depend only on $|\rho|$ since
Spin$(s,t)$ = Spin$(t,s)$.

The dimension $D$ mod(8) determines the nature of the
Spin($V$)-morphisms of the spinor representation $S$. Let $g\in
{\rm Spin}(V)$ and let  $\Sigma(g):S\longrightarrow S$  and
$L(g):V\longrightarrow V$ the spinor and vector representations of
$l\in{\rm Spin}(V)$ respectively. Then a
 map $A$
$$ A: S\otimes S
\longrightarrow \Lambda^k ,$$
 where $\Lambda^k=\Lambda^k(V)$ are
the $k$-forms on $V$, is a Spin($V$)-morphism if
$$A(\Sigma(g)s_1\otimes \Sigma(g)s_2)=L^k(g)A(s_1\otimes s_2). $$

\medskip

In Tables \ref{realprop} and \ref{morphisms}, reality and symmetry properties of spinors are reported.

\begin{table}[ht]
\begin{center}
\begin{tabular} {|c |c| c||c|c|c|}
\hline $\rho_0$(odd) &real dim($S$) & reality &$\rho_0$(even)
&real dim($S^{\pm}$) & reality\\ \hline \hline 1& $2^{(D-1)/2}$
&$\Bbb R$ & 0 &$2^{D/2-1}$&$\Bbb R$ \\ \hline 3& $2^{(D+1)/2}$ &$\Bbb H$& 2
&$2^{D/2}$&$\Bbb C$ \\ \hline 5& $2^{(D+1)/2}$ &$\Bbb H$& 4
&$2^{D/2}$&$\Bbb H$\\ \hline 7& $2^{(D-1)/2}$ &$\Bbb R$ & 6
&$2^{D/2}$&$\Bbb C$ \\ \hline
\end{tabular}
\caption{Reality properties of spinors}\label{realprop}
\end{center}
\end{table}

\bigskip
 
 \begin{table}[ht]
\begin{center}
\begin{tabular} {|c |c|c||c|c|}
\hline \multicolumn{1}{|c|}{$D$} &\multicolumn{2}{| c||}{$k $
even}&\multicolumn{2}{|c|}{$k $ odd}\\\hline\hline & morphism &
symmetry &morphism&symmetry\\
  \hline
  0& $S^\pm\otimes S^\pm\rightarrow\Lambda^k$&$(-1)^{k(k-1)/2}$& $S^\pm\otimes
  S^\mp\rightarrow\Lambda^k$&\\\hline
  1& $S\otimes S\rightarrow\Lambda^k$&$(-1)^{k(k-1)/2}$& $S\otimes
  S\rightarrow\Lambda^k$&$(-1)^{k(k-1)/2}$\\\hline
  2& $S^\pm\otimes S^\mp\rightarrow\Lambda^k$&& $S^\pm\otimes
  S^\pm\rightarrow\Lambda^k$&$(-1)^{k(k-1)/2}$\\\hline
  3& $S\otimes S\rightarrow\Lambda^k$&$-(-1)^{k(k-1)/2}$& $S\otimes
  S\rightarrow\Lambda^k$&$(-1)^{k(k-1)/2}$\\\hline
  4& $S^\pm\otimes S^\pm\rightarrow\Lambda^k$&$-(-1)^{k(k-1)/2}$& $S^\pm\otimes
  S^\mp\rightarrow\Lambda^k$&\\\hline
  5& $S\otimes S\rightarrow\Lambda^k$&$-(-1)^{k(k-1)/2}$& $S\otimes
  S\rightarrow\Lambda^k$&$-(-1)^{k(k-1)/2}$\\\hline
  6& $S^\pm\otimes S^\mp\rightarrow\Lambda^k$&& $S^\pm\otimes
  S^\pm\rightarrow\Lambda^k$&$-(-1)^{k(k-1)/2}$\\\hline
  7& $S\otimes S\rightarrow\Lambda^k$&$(-1)^{k(k-1)/2}$& $S\otimes
  S\rightarrow\Lambda^k$&$-(-1)^{k(k-1)/2}$\\\hline

\end{tabular}
\caption{Properties of morphisms.}\label{morphisms}
\end{center}
\end{table}

\section{Orthogonal, symplectic and linear spinors}

We consider now morphisms $$S\otimes S\longrightarrow
\Lambda^0\simeq \Bbb C.$$ If a morphism of this kind exists, it is
unique up to a multiplicative factor. The vector space of the
spinor representation has then a bilinear form invariant under
Spin($V$).
 Looking at Table \ref{morphisms}, one can see that this morphism
 exists except for $D_0=2,6$, where instead a morphism
 $$S^{\pm}\otimes S^{\mp}\longrightarrow \Bbb C$$
 occurs.

 We shall call a spinor representation orthogonal if it has a
 symmetric, invariant bilinear form. This happens for $D_0=0,1,7$ and
 Spin$(V)^{\Bbb C}$ (complexification of Spin($V$)) is then  a subgroup
 of the complex orthogonal group ${\rm SO}(n,\Bbb C)$, where $n$ is the
 dimension of the spinor representation (Weyl spinors for $D$ even).
 The generators
 of SO$(n,\Bbb C)$ are  $n\times n$ antisymmetric matrices. These are obtained
 in terms of the morphisms
$$S\otimes S\longrightarrow\Lambda^k,$$ which are  antisymmetric.
This gives  the decomposition of the adjoint representation of
${\rm SO}(n,\Bbb C)$ under the subgroup ${\rm Spin}(V)^{\Bbb C}$.
 In particular, for $k=2$ one obtains the generators of ${\rm Spin}(V)^{\Bbb C}$.

 A spinor representation is called symplectic if it has an
 antisymmetric, invariant bilinear form. This is the case for
 $D_0=3,4,5$. ${\rm Spin}(V)^{\Bbb C}$ is a subgroup of the symplectic group
 ${\rm Sp}(2p,\Bbb C)$, where $2p$ is the dimension of the
 spinor representation. The Lie algebra ${\rm sp}(2p,\Bbb C)$ is formed by all the
 symmetric matrices, so it is  given in terms of the
  morphisms $S\otimes S\rightarrow \Lambda^k$ which are symmetric.
 The generators
 of ${\rm Spin}(V)^{\Bbb C}$ correspond to $k=2$ and  are symmetric matrices.

 For $D_0=2,6$
 one has an
 invariant morphism
 $$B:S^{+}\otimes S^{-}\longrightarrow \Bbb C.$$
The representations $S^+$ and $S^-$ are
one the contragradient (or dual) of the other.
 The spin representations extend to  representations of the
 linear group ${\rm GL}(n,\Bbb C)$, which leaves the pairing  $B$ invariant. These
 spinors are called linear. Spin($V)^{\Bbb C}$ is a subgroup of the simple factor 
 SL$(n,\Bbb C)$.

These properties depend exclusively on the dimension\cite{de}. When combined with the
reality properties, which depend on
$\rho$, one obtains real groups embedded in ${\rm SO}(n,\Bbb C)$, ${\rm
Sp}(2p,\Bbb C)$ and
${\rm GL}(n,\Bbb C)$ which have an action on the space of the real spinor
representation $S^\sigma$. The real groups contain as a subgroup
${\rm Spin}(V)$.

We need first some general facts about real forms of simple Lie
algebras\cite{de}. Let $S$ be a complex vector space of dimension $n$ which
carries an irreducible representation of a complex Lie algebra
$\g$. Let $G$ be the complex Lie group associated to $\g$.  Let
$\sigma$ be a conjugation or a  pseudoconjugation on $S$ such that
$\sigma X\sigma^{-1}\in \g$ for all $X \in \g.$ Then  the map
$$X\mapsto X^\sigma=\sigma X\sigma^{-1}$$ is a conjugation of
$\g$. We shall write $${\g}^\sigma=\{X\in \g|X^\sigma=X\}.$$
${\g}^\sigma$ is a real form of $\g$. If $\tau=h\sigma h^{-1}$,
with $h\in \g$, ${\g}^\tau=h{\g}^\sigma h^{-1}$. ${\g}^\sigma={\g}^{\sigma'}$
if and only if $\sigma'=\epsilon \sigma$ for $\epsilon$ a scalar with $|\epsilon|=1$;
in particular, if ${\g}^\sigma$ and ${\g}^\tau$ are conjugate by $G$,
$\sigma$ and $\tau$ are both conjugations or both pseudoconjugations. The conjugation can
also be defined on the group $G$, $g\mapsto \sigma g\sigma^{-1}$.

\section{Real forms of the classical Lie algebras}

We  describe  the real forms of the classical Lie algebras from this
point of view\cite{aaa}. (See also Ref.~\cite{he}).

\paragraph{Linear algebra, sl($\mathbf{ S}$).}

\subparagraph{(a)} If $\sigma$ is a conjugation of $S$, then
there is an isomorphism  $S\rightarrow  {\Bbb C}^n$  such that $\sigma$
goes over to the standard conjugation of ${\Bbb C}^n$. Then
${\g}^\sigma\simeq{\rm sl}(n,\Bbb R)$. (The conjugation acting  on gl$(n,\Bbb C)$ gives
 the real form gl($n,\Bbb R$)).

\subparagraph{(b)} If $\sigma$ is a pseudoconjugation and $\g$
doesn't leave invariant a non degenerate bilinear form, then there
is an isomorphism of $S$ with ${\Bbb C}^n$, $n=2p$ such that $\sigma$
goes over to $$(z_1,\dots,z_p,z_{p+1},\dots z_{2p})\mapsto
(z^*_{p+1},\dots z^*_{2p},-z^*_1,\dots,-z^*_p).$$ Then
${\g}^\sigma\simeq {\rm su}^*(2p)$. (The pseudoconjugation acting in
on gl$(2p,\Bbb C)$ gives
 the real form ${\rm su}^*(2p)\oplus{\rm so}(1,1)$.)

To see this, it is enough to prove that ${\g}^\sigma$ does not leave
invariant any non degenerate hermitian form, so it cannot be of
the type su$(p,q)$. Suppose that $\langle\cdot ,\cdot\rangle$ is a
${\g}^\sigma$-invariant, non degenerate hermitian form.  Define
$(s_1,s_2):=\langle\sigma (s_1),s_2\rangle$. Then $(\cdot ,\cdot)$
is bilinear and ${\g}^\sigma$-invariant, so it is also
$\g$-invariant.

\subparagraph{(c)} The remaining cases, following E. Cartan's
classification of real forms of simple Lie algebras, are
${\rm su}(p,q)$, where a non degenerate hermitian bilinear form is
left invariant. They do not correspond to a conjugation or
pseudoconjugation on $S$, the space of the fundamental
representation. (The real form of gl$(n,\Bbb C)$ is in this case u$(p,q)$).

\paragraph{Orthogonal algebra, so($\mathbf{S}$).} $\g$ leaves invariant a non degenerate,
symmetric bilinear form. We will denote it by $(\cdot,\cdot)$.

\subparagraph{(a)} If $\sigma$ is a conjugation preserving $\g$, one can prove that
 there is an isomorphism of
$S$ with ${\Bbb C}^n$ such that  $(\cdot,\cdot)$ goes to the standard
form and ${\g}^\sigma$ to ${\rm so}(p,q)$, $p+q=n$. Moreover, all
${\rm so}(p,q)$ are obtained in this form.

\subparagraph{(b)} If $\sigma$ is a pseudoconjugation preserving $\g$,
 ${\g}^\sigma$ cannot be
any of the ${\rm so}(p,q)$. By E. Cartan's classification, the only
other possibility is that ${\g}^\sigma\simeq {\rm so}^*(2p)$. There
is an isomorphism of $S$ with ${\Bbb C}^{2p}$ such that $\sigma$ goes to
$$(z_1,\dots z_p,z_{p+1},\dots z_{2p})\mapsto (z^*_{p+1},\dots
z^*_{2p},-z^*_{1},\dots -z^*_{p}).$$

\paragraph{Symplectic algebra, sp($\mathbf{S}$).} We denote by $(\cdot,\cdot)$ the symplectic
form on $S$. \subparagraph{(a)}If $\sigma$ is a conjugation
preserving $\g$, it is clear  that there is an isomorphism of $S$
with ${\Bbb C}^{2p}$, such that ${\g}^\sigma\simeq {\rm sp}(2p,\Bbb R)$.

\subparagraph{(b)}If $\sigma$ is a pseudoconjugation preserving
$\g$, then ${\g}^\sigma\simeq {\rm usp}(p,q)$, $p+q=n=2m, \; p=2p',\;
q=2q' $. All the real forms ${\rm usp}(p,q)$ arise in this way.
There is an isomorphism of $S$ with ${\Bbb C}^{2p}$ such that $\sigma$
goes to $$(z_1,\dots z_m,z_{m+1},\dots z_{n})\mapsto
J_mK_{p',q'}(z^*_1,\dots z^*_m,z^*_{m+1},\dots z^*_{n}),$$ where

$$
J_m=\pmatrix {0 & I_{m\times m}\cr
-I_{m\times m} & 0},\qquad
 K_{p',q'}=\pmatrix {-I_{p'\times p'}
&0&0&0\cr
0&I_{q'\times q'}&0&0\cr 
0&0&-I_{p'\times
p'}&0\cr
0&0&0&I_{q'\times q'}}.
$$

\bigskip

 In Section 2 we saw that there is  a conjugation on $S$
when  the spinors are real and a pseudoconjugation when they are
quaternionic\cite{aaa} (both denoted by $\sigma$). We have a group,
${\rm SO}(n,\Bbb C)$, ${\rm Sp}(2p,\Bbb C)$ or
 ${\rm GL}(n,\Bbb C)$ acting on $S$ and  containing ${\rm Spin}(V)^{\Bbb C}$.  We note
that this group is minimal in the classical group series. If the
Lie algebra $\g$ of this group is stable under the conjugation
$$X\mapsto \sigma X\sigma^{-1}$$ then the real Lie algebra
${\g}^\sigma$ acts on $S^\sigma$ and contains the Lie algebra of
${\rm Spin}(V)$. We shall call it the Spin($V$)-algebra.

Let $B$ be the space of ${\rm Spin}(V)^{\Bbb C}$-invariant bilinear forms
on $S$. Since the representation on $S$ is irreducible, this space
is at most one dimensional. Let it be one dimensional and let
 $\sigma$ be  a conjugation or a
pseudoconjugation and let $\psi\in B$. We define a conjugation
in the space $B$ as \begin{eqnarray*} B&\longrightarrow &B\\
\psi&\mapsto& \psi^\sigma\end{eqnarray*} $$
\psi^\sigma(v,u)=\psi(\sigma(v),\sigma(u))^*.$$ It is then
immediate that we can choose $\psi\in B$ such that
$\psi^\sigma=\psi$. Then if $X$ belongs to the Lie algebra
preserving $\psi$, so does $\sigma X\sigma^{-1}$.

One can determine   the real Lie algebras  in each
case\cite{aaa}. All the possible cases must be studied separately. 
 All dimension and signature relations are
mod(8). In the following, a relation like ${\rm Spin}(V)\subseteq
G$ for a group $G$ will mean  that the image of ${\rm Spin}(V)$
under the spinor representation  is in the connected component of
$G$. The same applies for the  relation ${\rm Spin}(V)\simeq G$.  For $\rho$ =
0,1,7 spin algebras commute with a conjugation, for $\rho$ = 3,4,5 they commute
with a pseudoconjugation.  For $\rho$ = 2,6 they are complex.  The complete
classification is reported in Table 3.

\begin{table}[ht]
\begin{center}
\begin{tabular} {|l |l|l|}
\hline 
Orthogonal&Real, $\rho_0=1,7$&${\rm so}(2^{\frac{(D-1)}{2}},\Bbb R)$ if $D=\rho$\\
\cline{3-3}
$D_0=1,7$&& ${\rm so}(2^{\frac{(D-1)}{2}-1},2^{\frac{(D-1)}{2}-1})$
if $D\neq\rho$\\
\cline{2-3} & Quaternionic,~$\rho_0=3,5$&${\rm so}^*(2^{\frac{(D-1)}{2}})$\\
\hline\hline 
Symplectic&Real,
$\rho_0=1,7$& ${\rm sp}(2^{\frac{(D-1)}{2}},\Bbb R)$\\
\cline{2-3}
$D_0=3,5$& Quaternionic,~$\rho_0=3,5$& ${\rm
usp}(2^{\frac{(D-1)}{2}},\Bbb R)$  if $D=\rho$\\
\cline{3-3} && ${\rm
usp}(2^{\frac{(D-1)}{2}-1},2^{\frac{(D-1)}{2}-1})$ if $D\neq
\rho$\\
\hline\hline\hline
 Orthogonal&Real, $\rho_0=0$&${\rm
so}(2^{\frac{D}{2}-1},\Bbb R)$ if $D=\rho$\\  \cline{3-3} $D_0=0$&&
${\rm so}(2^{\frac{D}{2}-2},2^{\frac{D}{2}-2})$
 if
$D\neq\rho$\\\cline{2-3}
 & Quaternionic,~$\rho_0=4$&${\rm
so}^*(2^{\frac{D}{2}-1})$\\\cline{2-3} &Complex, $\rho_0=2,6$&
${\rm so}(2^{\frac{D}{2}-1},{\Bbb C})_{\Bbb R}$\\
\hline\hline Symplectic&Real,
$\rho_0=0$& ${\rm sp}(2^{\frac{D}{2}-1},\Bbb R)$\\\cline{2-3} $D_0=4$&
Quaternionic,~$\rho_0=4$& ${\rm usp}(2^{\frac{D}{2}-1},\Bbb R)$ if
$D=\rho$\\
\cline{3-3} && ${\rm
usp}(2^{\frac{D}{2}-2},2^{\frac{D}{2}-2})$ if $D\neq
\rho$\\\cline{2-3} &Complex, $\rho_0=2,6$&${\rm
sp}(2^{\frac{D}{2}-1},{\Bbb C})_{\Bbb R}$\\
\hline\hline
Linear&Real,
$\rho_0=0$& ${\rm sl}(2^{\frac{D}{2}-1},\Bbb R)$\\
\cline{2-3}
$D_0=2,6$& Quaternionic,~$\rho_0=4$& ${\rm
su}^*(2^{\frac{D}{2}-1})$ \\\cline{2-3} &Complex, $\rho_0=2,6$&
${\rm su}(2^{\frac{D}{2}-1})$ if $D= \rho$\\
\cline{3-3} &&${\rm
su}(2^{\frac{D}{2}-2},2^{\frac{D}{2}-2})$ if $D\neq \rho$\\\hline
\end{tabular}
\caption{Spin$(s,t)$ algebras.}\label{spinalgebra}
\end{center}
\end{table}

\section{ Spin$\mathbf{(V)}$ superalgebras}

 We now consider the embedding of ${\rm Spin}(V)$ in simple real superalgebras.
We require in general  that the odd generators are in a real
  spinor representation of
  ${\rm Spin}(V)$. In the cases $D_0=2,6$, $\rho_0=0,4$ we have to allow the two independent
irreducible  representations, $S^+$ and $S^-$ in the superalgebra, since  the
relevant morphism is
$$S^+\otimes S^-\longrightarrow \Lambda^2.$$
The algebra is then non chiral.

We first consider minimal superalgebras\cite{na,vp} i.e. those with the
  minimal even subalgebra. From the classification of simple superalgebras
  \cite{nrs,ka,mp} one
obtains the results listed in Table~4.

\begin{table}
\begin{center}
\begin{tabular} {|c|c|l|l|}
\hline
 $D_0$   & $\rho_0$& Spin($V$) algebra&Spin($V$) superalgebra\\\hline\hline
 1,7& 1,7& so($2^{(D-3)/2},2^{(D-3)/2}$)&  \\\hline
 1,7& 3,5 & so$^*$($2^{(D-1)/2}$)&osp$(2^{(D-1)/2})^*|2)$ \\\hline
 3,5& 1,7& sp($2^{(D-1)/2},\Bbb R$)&osp$(1|2^{(D-1)/2},\Bbb R)$\\\hline
 3,5& 3,5& usp($2^{(D-3)/2},2^{(D-3)/2}$)&
 \\\hline\hline
 0& 0&so($2^{(D-4)/2},2^{(D-4)/2}$) &  \\\hline
 0& 2,6&so($2^{(D-2)/2},{\Bbb C})^{\Bbb R}$ &  \\\hline
 0& 4 & so$^*(2^{(D-2)/2}$)&osp$(2^{(D-2)/2})^*|2)$\\\hline
 2,6&  0&sl$(2^{(D-2)/2},\Bbb R)$&sl$(2^{(D-2)/2}|1)$ \\\hline
 2,6&2,6& su$(2^{(D-4)/2},2^{(D-4)/2})$&su$(2^{(D-4)/2},2^{(D-4)/2}|1)$\\\hline
2,6 & 4 &su$^*(2^{(D-2)/2}))$&su$(2^{(D-2)/2})^*|2)$ \\\hline
  4&0&sp($2^{(D-2)/2},\Bbb R$)&osp$(1|2^{(D-2)/2},\Bbb R)$ \\\hline
 4&2,6&sp($2^{(D-2)/2},{\Bbb C})^{\Bbb R}$&osp$(1|2^{(D-2)/2},\Bbb C)$ \\\hline
 4&4&usp($2^{(D-4)/2},2^{(D-4)/2}$)&  \\\hline
\end{tabular}
\caption{Minimal Spin($V$) superalgebras.}\label{min}
\end{center}
\end{table}

We note that the even part of the minimal superalgebra contains the Spin($V$) algebra
 obtained in Section~4 as a simple factor. For all quaternionic cases,
 $\rho_0=3,4,5$, a second simple factor SU(2) is present. For the linear cases there is
 an additional
non simple factor, SO(1,1) or U(1), as discussed in Section~4.

For $D=7$ and $\rho=3$ there is actually a smaller
 superalgebra, the exceptional  superalgebra  $f(4)$ with
 bosonic part  spin(5,2)$\times$su(2). The superalgebra
appearing in  Table \ref{min} belongs to the classical series and
its  even part is  so$^*(8)\times$su(2), being so$^*(8)$ the
Spin$(5,2)$-algebra.

Keeping the same number of odd generators, the  maximal simple  superalgebra
 containing ${\rm Spin}(V)$ is  an
 orthosymplectic algebra with ${\rm Spin}(V)\subset {\rm Sp}(2n,\Bbb R)$, being $2n$ the real
dimension of $S$. The various cases  are displayed in the Table~5. We note that the minimal superalgebra is not a
subalgebra of the maximal one, although it is so for the bosonic
parts.

\begin{table}
\begin{center}
\begin{tabular} {|c|c|l|}
\hline
 $D_0$   & $\rho_0$& Orthosymplectic\\\hline\hline
3,5,& 1,7& osp$(1|2^{(D-1)/2},\Bbb R)$ \\\hline
 1,7& 3,5  & osp$(1|2^{(D+1)/2},\Bbb R)$ \\\hline
  0&4 &osp$(1|2^{D/2},\Bbb R)$  \\\hline 4&0& osp$(1|2^{(D-2)/2},\Bbb R)$
\\\hline
 4&2,6  &osp$(1|2^{D/2},\Bbb R)$ \\\hline
  2,6&  0&osp$(1|2^{D/2},\Bbb R)$ \\\hline
  2,6 & 4 & osp$(1|2^{(D+2)/2},\Bbb R)$\\\hline
 2,6&2,6&osp$(1|2^{D/2},\Bbb R)$ \\\hline
\end{tabular}
\caption{Orthosymplectic Spin($V$) superalgebras}\label{max}
\end{center}
\end{table}

\section{Extended Superalgebras}
The present analysis can be generalized to the case of $N$ copies of the spinor
representation of spin $(s,t)$-algebras\cite{bb}.  By looking at the classification of
classical simple superalgebras\cite{na}--\cite{bg}, we find extensions for all $N$, where the number
of supersymmetries is always even if spinors are quaternionic (because of reality
properties) or orthogonal (because of symmetry properties).

In Table~6 the classification analogous to the one in Table~4 is given. 
SuperPoincar\'e algebras can be obtained from the simple superalgebras either by
contraction Spin$(s,t) \rightarrow$ InSpin$(s,t-1)$ or as subalgebras Spin$(s,t)
\rightarrow$ InSpin$(s-1,t-1)$. It is important to observe that the
$R$-symmetry may be non-compact for different signatures of space-time.

In fact the conjugation properties of the $R$-symmetry algebra is the same of
the space-time part.

As an example if we consider Euclidean four-dimensional $N = 2$ and $N = 4$
Yang-Mills theory, the $R$-symmetry becomes respectively SU(2) $\times$ SO(1,1)
and SU*(4).  The first case was considered long ago by
Zumino\cite{bz}.  These are the superalgebras appropriate for Yang-Mills instantons.  On
the other hand, if we consider a Minkowskian space with signature (2,2) the
$R$-symmetry is $GL(2,{\Bbb R})$ (for $N = 2$) and $SL(4,{\Bbb R})$ for $N = 4$.

Compact $R$-symmetries occur for $q = 0$ in Table~6, including all cases when the conformal
group SO(D,2) corresponds to ordinary Minkowski space $V_{(D-1,1)}$.

 \begin{table}
\begin{center}
\begin{tabular} {c|c|c|l |l|}

\cline{2-5} &$D_0$&$\rho_0$& R-symmetry&Spin$(s,t)$
superalgebra\\\cline{2-5} & 1,7& 1,7& ${\rm
sp}(2N,\Bbb R)$&${\rm
osp}(2^{\frac{D-3}{2}},2^{\frac{D-3}{2}}|2N,\Bbb R)$\\\cline{2-5}
& 1,7& 3,5& ${\rm usp}(2N-2q,2q)$&${\rm
osp}(2^{\frac{D-1}{2}\,*}|2N-2q,2q)$\\\cline{2-5} & 3,5&
1,7& ${\rm so}(N-q,q)$&${\rm
osp}(N-q,q|2^{\frac{D-1}{2}})$\\\cline{2-5} & 3,5& 3,5&
${\rm so}^*(2N)$&${\rm
osp}(2{N}^*|2^{\frac{D-3}{2}},2^{\frac{D-3}{2}})$\\\cline{2-5}
\cline{2-5} & 0& 0& ${\rm sp}(2N,\Bbb R)$&${\rm
osp}(2^{\frac{D-4}{2}},2^{\frac{D-4}{2}}|2N)$\\\cline{2-5}
& 0& 2,6& ${\rm sp}(2N,{\Bbb C})_{\Bbb R}$&${\rm
osp}(2^{\frac{D-2}{2}}|2N,{\Bbb C})_{\Bbb R}$\\\cline{2-5} & 0& 4&
${\rm usp}(2N-2q,2q)$&${\rm osp}(2^{\frac{D-2}{2}\,*}|2N-2q,2q)$\\\cline{2-5}
& 2,6& 0& ${\rm sl}(N,\Bbb R)$&${\rm
sl}(2^{\frac{D-2}{2}}|N,\Bbb R)$\\\cline{2-5} &2,6& 2,6& ${\rm
su}(N-q,q)$&${\rm
su}(2^{\frac{D-4}{2}},2^{\frac{D-4}{2}}|N-q,q)$\\
\cline{2-5}
& 2,6& 4& ${\rm su}^*(2N,\Bbb R)$&${\rm
su}(2^{\frac{D-2}{2\,*}}|2N^*)$\\\cline{2-5} & 4& 0& ${\rm
so}(N-q,q)$&${\rm osp}(N-q,q|2^{\frac{D-2}{2}})$\\
\cline{2-5} & 4&
2,6& ${\rm so}(N,{\Bbb C})_{\Bbb R}$&${\rm
osp}(N|2^{\frac{D-2}{2}},{\Bbb C})_{\Bbb R}$\\
\cline{2-5} & 4& 4& ${\rm
so}^*(2N)$&${\rm
osp}(2{N}^*|2^{\frac{D-4}{2}},2^{\frac{D-4}{2}})$\\\cline{2-5}
\end{tabular}
\caption{$N$-extended Spin$(s,t)$ superalgebras.}\label{next}
\end{center}
\end{table}
\vfill\eject
\vskip0.5cm
\noindent
{\large \bf Acknowledgements}

This lecture is based on collaborations with R. D'Auria and M.A. Ll\'edo and
R.~Varadarajan.  Enlightning conversations with A. van Proeyen and R. Stora are also
acknowledged.
This work has been supported in part by the European Commission RTN network
HPRN-CT-2000-00131, (Laboratori Nazionali di Frascati, INFN) and by the D.O.E. grant
DE-FG03-91ER40662, Task C.


\end{document}